\documentclass[sigconf, screen, nonacm, dvipsnames]{acmart}

\usepackage[noend]{algorithmic}
\usepackage{algorithm}
\usepackage{multirow}
\usepackage{amsmath}
\usepackage{bbm}
\usepackage{booktabs}
\usepackage[inline]{enumitem} 
\usepackage{subcaption}
\usepackage{bm} 
\usepackage{xcolor}

\usepackage{hyperref}
\usepackage{lipsum}

\usepackage{wrapfig}

\usepackage{arydshln}
\makeatletter
\def\adl@drawiv#1#2#3{
        \hskip.5\tabcolsep
        \xleaders#3{#2.5\@tempdimb #1{1}#2.5\@tempdimb}%
                #2\z@ plus1fil minus1fil\relax
        \hskip.5\tabcolsep}
\newcommand{\cdashlinelr}[1]{%
  \noalign{\vskip\aboverulesep
          \global\let\@dashdrawstore\adl@draw
          \global\let\ adl@draw\adl@drawiv}
  \cdashline{#1}
  \noalign{\global\let\adl@draw\@dashdrawstore
          \vskip\belowrulesep}}
\makeatother

\usepackage{tikz}
\usetikzlibrary{automata, arrows, bayesnet, bending}

\usepackage{tikzscale}

\copyrightyear{2024}
\acmYear{2024}
\setcopyright{none}
\acmConference[RecSys '24]{Proceedings of the 18th ACM Conference on Recommender Systems}{October 14--18, 2024}{Bari, Italy}
\acmBooktitle{Proceedings of the 18th ACM Conference on Recommender Systems (RecSys '24), October 14--18, 2024, Bari, Italy}

\begin{document}

\title{Powerful A/B-Testing Metrics and Where to Find Them}

\author{Olivier Jeunen}
\affiliation{
  \institution{ShareChat}
  \city{Edinburgh}
  \country{United Kingdom}
}

\author{Shubham Baweja}
\affiliation{
  \institution{ShareChat}
  \city{Bangalore}
  \country{India}
}

\author{Neeti Pokharna}
\affiliation{
  \institution{ShareChat}
  \city{Bangalore}
  \country{India}
}

\author{Aleksei Ustimenko}
\affiliation{
  \institution{ShareChat}
  \city{London}
  \country{United Kingdom}
}

\begin{abstract}
Online controlled experiments, colloquially known as A/B-tests, are the bread and butter of real-world recommender system evaluation.
Typically, end-users are randomly assigned some system variant, and a plethora of metrics are then tracked, collected, and aggregated throughout the experiment.
A North Star metric (e.g. long-term growth or revenue) is used to assess which system variant should be deemed superior.
As a result, most collected metrics are \emph{supporting} in nature, and serve to either (i) provide an understanding of how the experiment impacts user experience, or (ii) allow for confident decision-making when the North Star metric moves insignificantly (i.e. a false negative or type-II error).
The latter is not straightforward: suppose a treatment variant leads to fewer but longer sessions, with more views but fewer engagements; should this be considered a positive or negative outcome?

The question then becomes: how do we assess a supporting metric's utility when it comes to decision-making using A/B-testing?
Online platforms typically run dozens of experiments at any given time.
This provides a wealth of information about interventions and treatment effects that can be used to evaluate metrics' utility for online evaluation.
We propose to collect this information and leverage it to quantify type-I, type-II, and type-III errors for the metrics of interest, alongside a distribution of measurements of their statistical power (e.g. $z$-scores and $p$-values).
We present results and insights from building this pipeline at scale for two large-scale short-video platforms: ShareChat and Moj; leveraging hundreds of past experiments to find online metrics with high statistical power.
\end{abstract}

\maketitle

\section{Introduction \& Motivation}
Continuous experimentation practices have been at the heart of progress for modern platforms on the web~\cite{kohavi2020trustworthy}.
By defining a North Star metric (e.g. long-term growth or revenue) and following rigorous statistical procedures as popularised by \citet{Fisher1921} and \citet{Rubin1974}, causal effect estimates of system changes on that North Star metric are used to guide system development and progression.\footnote{It is worth noting that there are common pitfalls to be avoided~\cite{Dmitriev2017,Kohavi2022,Jeunen2023_Forum}.}
Many experiments fail to move the North Star metric significantly, which can in part be attributed to the low \textit{sensitivity} of the metric.
As a result, a large body of research originating from industry focuses on the development of methods that reduce the variance of the causal effect estimates we obtain from A/B-tests~\cite{Deng2013,Poyarkov2016,Budylin2018,Baweja2024}. 
This, in turn, reduces the number of false negatives and increases the experimental velocity of the platform.

Naturally, many metrics are measured alongside the North Star.
Such metrics are typically supporting in nature, and many serve to guide decision-making when North Star measurements are statistically insignificant.
They are often referred to as \textit{proxy} metrics, and several recent works propose ways to identify short-term proxies that align with a longer-term North Star~\cite{Tripuraneni2023,Richardson2023}.
This line of work can be taken one step further, where ``optimal'' proxy metrics are \emph{learnt} from a dataset of past experiments that ran on the platform~\cite{Deng2016,Kharitonov2017,Tripuraneni2023,Jeunen2024_Learning}, for varying notions of ``optimality''.

In practice, A/B-testing results often allow for conflicting interpretations.
In the example of ShareChat, a large-scale short-video platform, we might observe experiments where the treatment variant leads to: fewer but longer sessions, more views but fewer engagements, slightly more skips but slightly more shares.
The question of whether this should be seen as a successful experiment, then becomes paramount to give guidance for the platform's evolution.
We propose to leverage the wealth of information that past experiments on the platform harness, to quantify type-I, type-II, and type-III errors for the metrics of interest~\cite{Urbano2019}, alongside a distribution of measurements of their statistical power (e.g. $z$-scores and $p$-values).
Such analyses unlock insights that are crucial to evaluate online evaluation metrics themselves, and benefits confident decision-making from online experiments going forward.
\begin{figure*}[t]
    \centering
    \vspace{-2ex}
    \includegraphics[width=\linewidth]{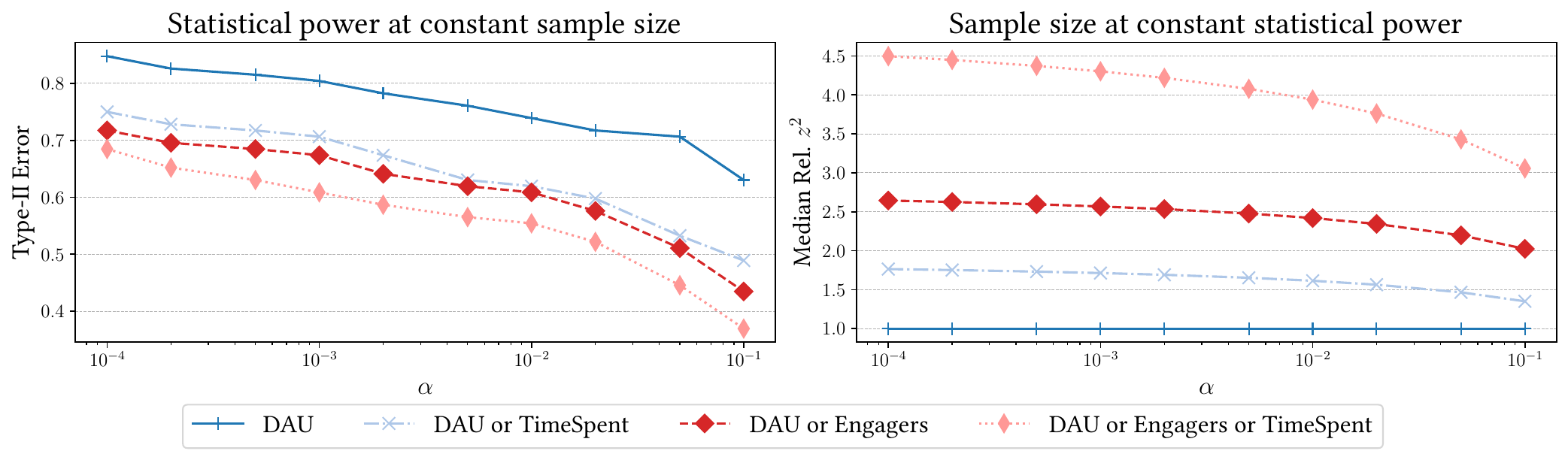}
    \caption{Empirical evaluation of various online evaluation metrics on ShareChat data: at a 95\% confidence level, we can reduce type-II errors by relative $35\%$ or reduce the necessary sample size by a factor $\times~3.5$ when considering the right set of metrics.}
    \label{fig:RQ}
    \vspace{-2ex}
\end{figure*}
\section{Methodology \& Contributions}
Our mathematical notation largely follows that of \citet{Jeunen2024_Learning}.
Whilst their work focuses on learning new metrics, ours is on building a pipeline that validates existing metrics.
When collecting past A/B experiments on the platform, we can divide them into three classes.
That is, for experiments with pairs of variants $\mathcal{E} = \{(A,B)_{i}\}_{i=1}^{|\mathcal{E}|}$, there are three distinct relations we can distinguish between pairs of deployed system variants:
\begin{enumerate}
    \item \emph{Known} outcomes: $(A,B) \in\mathcal{E}^{+}$, where $A \succ B$, 
    \item \emph{Unknown} outcomes: $(A,B) \in\mathcal{E}^{?}$, where $A~?~B$,  
    \item \emph{A/A} outcomes: $(A,B) \in\mathcal{E}^{\simeq}$, where $A \simeq B$.
\end{enumerate}
A \emph{known} and vetted preference of variant $A$ over $B$ --- typically because the North star or other guardrail metrics showed statistically significant improvements, is denoted by $A\succ B$. These insights have been replicated and validated to ensure trustworthiness.
\emph{Inconclusive} outcomes as denoted by $A~?~B$ imply statistically insignificant outcomes on the North Star.
Finally, $A \simeq B$ represents A/A experiments, where we know the null hypothesis to hold true (by design).
This dataset of past A/B experiments is collected and labelled by hand, from natural experiments occurring on the platform over time.
A/A experiments are sampled synthetically.
We measure key properties of online evaluation metrics in an experiment, such as:

\textbf{$z$-scores}, measuring how many standard deviations the null hypothesis effect size of $0$ is below the empirical mean:
$$ z_{m}^{A\succ B} = \frac{\mu_{m}^{A} - \mu_{m}^{B}}{\sqrt{\sigma^{A}_{m} + \sigma^{B}_{m}}}.
$$
The subscript $m$ denotes the metric, superscripts denote variants, $\mu$ is the sample mean and $\sigma$ the variance of the sample mean.

\textbf{Type-I errors} or \textit{false positives}, from A/A experiments:
$$ \varepsilon_{\rm I}(m) =
\frac{\left|\left\{ A,B \in \mathcal{E}^{\simeq} : \left|z_{m}^{A \succ B}\right| > \Phi^{-1}\left(1-\frac{\alpha}{2} \right) \right\} \right|}{\left|\mathcal{E}^{\simeq}\right|},
$$
where $\Phi^{-1}$ represents the standard Gaussian quantile function, used to transform $z$-scores into $p$-values for hypothesis testing.

\textbf{Type-II errors} or \textit{false negatives}, from A/B experiments:
$$ \varepsilon_{\rm II}(m) = 
\frac{\left|\left\{ A,B \in \mathcal{E}^{+}\cup\mathcal{E}^{?}  :  \left|z_{m}^{A \succ B}\right| < \Phi^{-1}\left(1-\frac{\alpha}{2} \right) \right\} \right|}{\left|\mathcal{E}^{+}\cup\mathcal{E}^{?}\right|}.
$$
Naturally, it is important here to ensure that the inconclusive experiments $\mathcal{E}^{?}$ are \emph{real} false negatives, and not just small system changes that do not affect outcomes significantly.
For an experiment to be included in this set, the change has to be of significant magnitude (e.g. deployment of a new model architecture).

\textbf{Type-III} errors or \textit{sign errors} measured from known outcomes:
$$ \varepsilon_{\rm III}(m) = 
\frac{\left|\left\{ A,B \in \mathcal{E}^{+} :  z_{m}^{A \succ B} < - \Phi^{-1}\left(1-\frac{\alpha}{2} \right) \right\} \right|}{\left|\mathcal{E}^{+}\right|}.
$$

\section{Empirical Results \& Discussion}
We can leverage the collected information as described above in many ways.
Here, we answer two example research questions that hold significant implications for the business:
\begin{description}
    \item[\textbf{RQ1}] \textit{Can we quantify the reduction in type-II errors we can obtain by combining multiple metrics to make decisions?} 
    \item[\textbf{RQ2}] \textit{Can we quantify the reduction in sample size we can obtain by combining multiple metrics to make decisions?} 
\end{description}
We consider three often-used proxy metrics, for which we have validated that no type-III (sign) errors occur: they do not disagree with the North Star metric in any of our past experiments.
These metrics are: Daily Active Users (\textbf{DAU}), a count of users and days with at least one positive engagement (\textbf{Engagers}), and the time users spend on the platform (\textbf{TimeSpent}).
We can use this \textit{set} of metrics for decision-making purposes: if any of the metrics changes statistically significantly, we declare the experiment as a positive outcome.
Naturally, we need to resort to multiple hypothesis testing corrections when doing this, to ensure type-I errors are kept in check.
We report results for this setup on a set of past A/B experiments at ShareChat, where we use a Bonferroni correction to leverage sets of metrics instead of single metrics.
Results are visualised in Figure~\ref{fig:RQ}, with type-II errors ($1-$ statistical power) on the left-hand side, and the median relative square $z$-score on the right-hand side.
The former provides an estimate of the number of false negative experiments we would have without changing the sample size, whereas the latter estimates the reduction in the sample size required to obtain constant statistical power to the standalone DAU metric.
We do not visualise type-I errors, but have empirically validated that the Bonferroni correction successfully alleviates them to the required significance level.
By using these sets of metrics, we can effectively increase the statistical power of our evaluation procedure (leading to more decisions being made confidently), or decrease the sample size that is required to obtain constant statistical power (equating to a decrease in the cost of experimentation for the platform).
\section{Conclusions \& Outlook}
A/B-tests are a staple in industry recommender systems, and widely regarded as the standard for online evaluation.
The practice of running A/B-tests not only provides valuable information about the treatments that are being tested, but also about the metrics that are being collected. 
Our work highlights how this data can be used to obtain valuable insights for platforms on the web, benefiting their experimental velocity by providing a pipeline that validates key desiderata for online evaluation metrics, such as statistical power.

\section*{Presenter Biography}
Olivier Jeunen is a Lead Applied Scientist at ShareChat with a PhD from the University of Antwerp, who has previously held positions at Amazon, Spotify, Facebook and Criteo.
His research focuses on applying ideas from causal and counterfactual inference to recommendation and advertising problems, which have led to 40+ peer reviewed contributions, two of which have been recognised with best paper awards. He is an active Program Committee member and reviewer for several conferences, journals and workshops---which has led to three outstanding reviewer awards at RecSys. Olivier is also an organising committee member at RecSys '22--'24 and ECIR '24, and routinely co-organises workshops and tutorials.

\bibliographystyle{ACM-Reference-Format}
\bibliography{bibliography}
\end{document}